\documentclass[preprint,aps,showpacs,pre]{revtex4}
\usepackage[dvips]{graphicx}
\usepackage{graphicx} 

\begin{document}
\title{Diffusion phenomenon in the hyperbolic and parabolic regimes}
\author{A. Sapora$^{1,2}$, M. Codegone $^{1,3}$, and G. Barbero$^{1,4}$}
\affiliation{$^1$Turin Polytechnical University in Tashkent,
 17, Niyazov Str. Sobir Rakhimov district Tashkent, 100095 Uzbekistan.\\
$^2$ Department of Structural, Building and Geotechnical Engineering\\
Politecnico di Torino Corso Duca degli Abruzzi 24
10129 Torino - Italy \\
$^3$ Department of Mathematical Sciences,
Politecnico di Torino, Corso Duca degli Abruzzi 24, 10129 Torino, Italy.\\
$^4$Department of Applied Science and Technology,
Politecnico di Torino, Corso Duca degli Abruzzi 24, 10129 Torino, Italy.
}
\begin{abstract}
We discuss the diffusion phenomenon in the parabolic and hyperbolic regimes. New effects related to the finite velocity of the diffusion process are predicted, that can partially explain the strange behavior associated to adsorption phenomenon. For sake of simplicity, the analysis is performed by considering a sample in the shape of a slab limited by two perfectly blocking surfaces, in such a manner that the problem is one dimensional in the space. Two cases are investigated. In the former, the initial distribution of the diffusing particles is assumed of gaussian type, centered around the symmetry surface in the middle of the sample. In the latter, the initial distribution is localized close to the limiting surfaces. In both cases, we show that the evolution toward to the equilibrium distribution is not monotonic. In particular, close to the limiting surfaces the bulk density of diffusing particles present maxima and minima related to the finite velocity of the diffusion process connected to the second order time derivative in the partial differential equation describing the evolution of the bulk density in the sample.
\end{abstract}
\date{\today}
\maketitle
\section{Introduction}
The fundamental relationship describing the diffusion of particles in an isotropic medium, the continuum approximation, is based on the diffusion equation, which is of parabolic type. This equation represents the conservation of the particles, and it has been written by assuming the validity of the law of Fick relating the current density to the gradient of concentration at the same time. Several papers have been devoted to generalize the diffusion equation \cite{L1,L2,L3}, mainly for its importance in the analysis of the experimental data relevant to the impedance spectroscopy technique. Our aim is to extend the standard model to the case in which subdiffusion occurs. We will base our analysis on the extended thermodynamics, where the diffusion current at the time $t$ depends on the gradient of concentration at the time $t-\tau$, where $\tau$ is a phenomenological parameter of the model. Our paper is organized as follows. The derivation of the parabolic and hyperbolic equations for the diffusion phenomenon is presented in Sect.II. The diffusion phenomenon in the parabolic and hyperbolic approximations are discussed in Sect.III and Sect.IV, respectively. The comparison between the predictions of the two models is presented in Sect.V. In that section we show that, due to the finite velocity of propagation of the bulk density variation, a non-monotonic time-dependence of the density of diffusing particles, in a given point of the sample, is observed.  In Sect.VI the problem is analyzed by means of the Wentzel-Kramers-Brillouin (WKB) method, to find an approximated solution taking into account the finite velocity of propagation of the bulk density variations, reducing to the parabolic solution when this velocity tends to infinite. The main points of our analysis are summarized in Sect.VII, devoted to the conclusions.

\section{Parabolic and Hyperbolic Approximations}
According the the law of Fick, if the bulk distribution of particles, $n$, is not homogeneous, a net density of current, ${\bf j}$, exists. The equation relating ${\bf j}$ with the spatial inhomogeneity of $n$ is
\begin{equation}
\label{01}{\bf j}({\bf r},t)=-D \nabla n({\bf r},t),
\end{equation}
where $D$ is the diffusion coefficient.
This equation states that the current density ${\bf j}({\bf r},t)$ at the point ${\bf r}$ and time $t$ is due to the gradient of the bulk density $ n({\bf r},t)$, at the same time $t$. By substituting Eq.(\ref{01}) in the equation of continuity
\begin{equation}
\label{02}\frac{\partial n({\bf r},t)}{\partial t}=-\nabla\cdot {\bf j}({\bf r},t)
\end{equation}
stating the conservation of the number of particles, we get
\begin{equation}
\label{03}\frac{\partial n({\bf r},t)}{\partial t}=D\,\nabla^2 n({\bf r},t)\,
\end{equation}
which is the diffusion equation. It has to be solved with the boundary condition
\begin{equation}
\label{04}{\bf k}\cdot {\bf j}({\bf r},t)=0,
\end{equation}
where ${\bf k}$ is the geometrical normal of the surface $S$ limiting the sample, outward directed. Equation (\ref{04}) holds in the case where the particles cannot leave the sample.  Equation (\ref{03}) with the boundary condition (\ref{04}) has to be solved with the initial condition $n({\bf r},0)=n_0({\bf r})$ \cite{cussler}. For a unbounded space the solution of Eq.(\ref{03}), in an isotropic medium, is \cite{smirnov}
\begin{equation}
\label{fick} n({\bf r},t)=\frac{1}{(4 \pi D t)^{3/2}}\,\int_{V_{\infty}}\,d{\bf u}\,n_0({\bf u})\,\exp\left\{-\frac{({\bf r}-{\bf u})^2}{4 Dt}\right\},
\end{equation}
where $d{\bf u}=d u_x\,du_y\,du_z$ and the integration on $V_{\infty}$ means on $-\infty \leq u_x \leq \infty$, $-\infty \leq u_y \leq \infty$, $-\infty \leq u_z \leq \infty$.

From Eq.(\ref{fick}) it follows that in the case of an unbounded domain, if the initial condition on the bulk distribution of the particles is of the type
\begin{equation}
\label{13-20}n({\bf r},0)={\cal{N}}\delta({\bf r}-{\bf r_0}),
\end{equation}
where ${\cal{N}}$ is the number of particles and $\delta({\bf r}-{\bf r_0})$ is the function of Dirac centered around ${\bf r_0}$, the bulk density of particles is given, for $t\geq 0$, by
\begin{equation}
\label{140}n({\bf r},t)=\frac{{\cal{N}}}{(4 \pi D t)^{3/2}}\,\exp\left\{-\frac{({\bf r}-{\bf r_0})^2}{4 Dt}\right\}.
\end{equation}
For $t\to 0$, $n({\bf r},t)$ is different from zero in all points of the domain. This means that the velocity of the bulk variation of density is infinite. Since this result follows from the integration of the diffusion equation, consequence of Eq.(\ref{01}), one can conclude that the law of Fick is an approximation for the diffusion current. As well known, if one faces the diffusion problem by means of the transport equation of Boltzmann, this absurd result is absent. Based on physical arguments, Cattaneo \cite{cattaneo} proposed to modify the law of Fick. The phenomenological derivation of the equation of Cattaneo in the case of diffusion is based on the assumption that the flux of particles ${\bf j}({\bf r},t)$ is given by the equation
\begin{equation}
\label{150}{\bf j}({\bf r},t)+\tau_r\,\frac{\partial {\bf j}({\bf r},t)}{\partial t}=-D\nabla n({\bf r},t),
\end{equation}
where $\tau_r$ is a positive parameter having the dimensions of a time. For $\tau_r=0$ we recover the law of Fick. Equation (\ref{150}) can be considered as an approximation of the functional relation
\begin{equation}
\label{160}{\bf j}({\bf r},t+\tau_r)=-D\nabla n({\bf r},t),
\end{equation}
when $\tau_r$ is a small parameter \cite{compte,criado,ramos,lewa,macdonald}. From Eq.(\ref{150}) we obtain
\begin{equation}
\label{170}\nabla \cdot{\bf j}({\bf r},t)+\tau_r\,\nabla \cdot\left(\frac{\partial {\bf j}({\bf r},t)}{\partial t}\right)=-D\nabla^2 n({\bf r},t),
\end{equation}
that by inverting the order of the derivations in the second term can be rewritten as
\begin{equation}
\label{180}\nabla \cdot{\bf j}({\bf r},t)+\tau_r\,\frac{\partial }{\partial t}\left[\nabla \cdot{\bf j}({\bf r},t)\right]=-D\nabla^2 n({\bf r},t).
\end{equation}
By taking into account the equation of continuity, Eq.(\ref{02}), from Eq.(\ref{180}) we get
\begin{equation}
\label{190}\frac{\partial n({\bf r},t)}{\partial t}+\tau_r\,\frac{\partial^2 n({\bf r},t)}{\partial t^2}=D\,\nabla^2 n({\bf r},t)\,,
\end{equation}
which is the generalization proposed by Cattaneo for the diffusion equation \cite{cattaneo}. From Eq.(\ref{190}) the velocity of propagation of the time variation of the bulk density variation is finite and given by $C=\sqrt{D/\tau_r}$.

\section{Diffusion phenomenon in the parabolic approximation}
We are interested in the evolution of an initial distribution of particle in an isotropic liquid. For sake of simplicity we assume that the sample is in the shape of a slab of thickness $d$, and that the limiting surfaces are completely blocking. We use a cartesian reference frame having the $z$-axis perpendicular to the liming surfaces, at $z=\pm d/2$. In this frame work the bulk density of diffusing particles is $n=n(z,t)$, and the initial distribution $n_0(z)=n(z,0)$. We will limit our analysis to the case where $n_0(z)=n_0(-z)$, which is rather important for practical application (the generalization to the case where the parity is not defined is straightforward).
In this case the bulk current density of diffusion has only the $z$-component, $j$, and as it follows from Eq.(\ref{04}) it vanishes on the limiting surfaces, i.e. $j(\pm d/2,t)=0$ for all $t$. In this simple case, Eq.(\ref{03}) can be rewritten as
\begin{equation}
\label{a-1}\frac{\partial n}{\partial t}=D\,\frac{\partial^2 n}{\partial z^2},
\end{equation}
that has to be solved with the boundary conditions
\begin{equation}
\label{a-2}-D\,\left( \frac{\partial n}{\partial z}\right)_{\pm d/2}=0.
\end{equation}
By introducing the dimensionless variables $u=z/d$ and $v=t/\tau_D$, where $\tau_D=d^2/D$ is the diffusion time, Eq.s(\ref{a-1},\ref{a-2}) can be  rewritten as
\begin{equation}
\label{a-3}\frac{\partial n}{\partial v}=\frac{\partial^2 n}{\partial u^2},
\end{equation}
and
\begin{equation}
\label{a-4}\left(\frac{\partial n}{\partial u}\right)_{\pm 1/2}=0,
\end{equation}
respectively. The initial condition for the problem under investigation is
\begin{equation}
\label{a-5} n(u,0)=n_0(u).
\end{equation}
As it is clear from Eq.(\ref{a-3}), for $v\to \infty$, $n$ tends to a constant, $n_{\rm eq}$, as expected. From the condition stating the conservation of particles it follows that
\begin{equation}
\label{a-6}n_{\rm eq}=\int_{-1/2}^{1/2} n_0(u)\,du.
\end{equation}
We look for a solution of Eq.(\ref{a-3}) of the type $n(u,v)=U(u) V(v)$. By substituting this ansatz into Eq.(\ref{a-3}) we get
\begin{equation}
\label{a-7}\frac{1}{V}\,\frac{d V}{dv}=\frac{1}{U}\,\frac{d^2 U}{du^2}=-\alpha^2,
\end{equation}
where $\alpha$ is a constant (separation constant) to be determined. By taking into account the linearity of the problem the solution we are looking for, for the assumed symmetry of the initial distribution $n_0(u)=n_0(-u)$, is
\begin{equation}
\label{a-8}n(u,v)=\sum_{\alpha}\, K_{\alpha}\, e^{-\alpha^2 v}\,\,\cos(\alpha u),
\end{equation}
where the coefficients $K_{\alpha}$ have to be determined by means of the initial condition (\ref{a-5}).
From Eq.(\ref{a-8}), by taking into account (\ref{a-2}), we get
\begin{equation}
\label{a-9}\left(\frac{\partial n}{\partial u}\right)_{\pm 1/2}=-\sum_{\alpha}\,\alpha \, K_{\alpha}\, e^{-\alpha^2 v}\,\,\sin\left(\frac{\alpha}{2}\right)=0,
\end{equation}
from which we obtain
\begin{equation}
\label{a-10}\alpha=2\,m\,\pi,
\end{equation}
where $m$ is an integer. Consequently, expansion (\ref{a-8}) can be rewritten as
\begin{equation}
\label{a-11}n(u,v)=K_0+\sum_{m=1}^{\infty} K_m\,e^{-(2m \pi)^2 v}\,\,\cos(2 m \pi u).
\end{equation}
The functions $\varphi_m(u)=\cos(2 \pi m u)$, for $m\geq 1$ are such that
\begin{equation}
\label{a-12}(\varphi_m,\varphi_k)=\int_{-1/2}^{1/2}\cos(2 m \pi u)\,\cos(2 k \pi u) \,du=\frac{1}{2}\,\delta_{mk},
\end{equation}
where $\delta_{mk}=1$ if $m=k$, and $\delta_{mk}=0$ for $m\neq k$. Thus, from the initial condition (\ref{a-5}), rewritten as
\begin{equation}
\label{a-13}n(u,0)=K_0+\sum_{m=1}^{\infty} K_m\,\cos(2 m \pi u),
\end{equation}
we get
\begin{eqnarray}
\label{a-14}K_0&=&\int_{-1/2}^{1/2}\, n(u,0)\,du=n_{\rm eq}\\
\label{a-15}
K_m&=&2\,\int_{-1/2}^{1/2}\,n(u,0)\,\cos(2 m \pi u)\,du,
\end{eqnarray}
for the coefficients entering into expansion (\ref{a-11}). The relations reported above are general, and hold all the time that $n_0(u)=n_0(-u)$. It is then possible to obtain the profile of the distribution $n(u,v)$ for each reduced time $v$ and in each point $u$. In the following we consider two particular cases of some importance from the experimental point of view.

The case in which the initial distribution of diffusing particles is given by
\begin{equation}
\label{a-16}n(u,0)=B\,\sqrt{\frac{b}{\pi}}\, e^{\displaystyle -b u^2},
\end{equation}
where $B$ is a normalization constant, corresponds to the situation where the particles are located for $t=0$ in the center of the sample in a region of the order $\ell\sim 1/b$. For $b\to \infty$, we get a delta Dirac distribution. In this framework using Eq.s(\ref{a-14},\ref{a-15}) we obtain for the expansion coefficients
\begin{eqnarray}
\label{a-17}K_0&=&B\,\,Erf\left(\sqrt\frac{b}{2}\right)\\
\label{a-18}K_m&=&B\,\left\{Erf\left(\frac{b-2 i m \pi}{2\sqrt{b}}\right)+Erf\left(\frac{b+2 i m \pi}{2\sqrt{b}}\right)\right\}\,\,\exp\left(-\frac{m^2 \pi^2}{b}\right),
\end{eqnarray}
where $Erf$ is the error function and $i$ the imaginary unit.

Another situation of some experimental importance is the one where the initial distribution of particles is localized close to the limiting surfaces. In this case
\begin{equation}
\label{a-19}n(u,0)=\frac{1}{2}\,A\,s\,\,\frac{\cosh(s u)}{\sinh(s/2)},
\end{equation}
where  $A$ is a normalization constant and $s$ a large number. For $s\to \infty$ we have that the initial distribution of diffusing particles is formed by two delta Dirac functions localized at the limiting surfaces. In this case the coefficients entering in the expansion (\ref{a-11}) are
\begin{eqnarray}
\label{a-20}K_0&=& A\\
\label{a-21}K_m&=&2 A  \,\,\frac{s^2\,(-1)^m}{(2 m \pi)^2+s^2}.
\end{eqnarray}
In the following, we will discuss the evolution of the distribution of particles obtained above, valid in the parabolic approximation of the diffusion equation, and compare them with the prediction of the hyperbolic approximation.

\section{Diffusion phenomenon in the hyperbolic approximation}
In the hyperbolic approximation the fundamental equation of the problem is Eq.(\ref{190}), that for our slab geometry reads
\begin{equation}
\label{b-1}\tau_r\,\frac{\partial^2 n}{\partial t^2}+\frac{\partial n}{\partial t}=D\,\frac{\partial^2 n}{\partial z^2}.
\end{equation}
In terms of dimensionless coordinates Eq.(\ref{b-1}) can be rewritten as
\begin{equation}
\label{b-2}\varepsilon\,\frac{\partial^2 n}{\partial v^2}+\frac{\partial n}{\partial v}=\frac{\partial^2 n}{\partial u^2},
\end{equation}
where $\varepsilon=\tau_r/\tau_D$ is a small quantity. Note that Eq.(\ref{b-2}) presents a singularity for $\varepsilon=0$, since the small parameter multiplies the higher derivative with respect to $v$ \cite{1,2}. A standard perturbational expansion of $n(u,v)$ in power of $\varepsilon$ is not possible \cite{3,4}. The standard approach WBK \cite{5,6} will be discussed in Sect.VI. We will solve Eq.(\ref{b-2}) using the separation of variables used in the parabolic case. For the present problem the initial conditions are the initial profile of the diffusing particles and the time derivative of such distribution. Hence Eq.(\ref{b-2}) has to be solved with the boundary condition (\ref{a-4}), related to the assumption that the limiting surfaces are blocking, and with the initial conditions
\begin{equation}
\label{b-3}n(u,0)=n_0(u),\quad{\rm and}\quad \left(\frac{\partial n}{\partial v}\right)_{v=0}=0.
\end{equation}
As before we assume that $n_0(u)=n_0(-u)$, since we are interested in the analysis of the cases (\ref{a-16},\ref{a-19}) considered above. By assuming, as in the previous case, that $n(u,v)=U(u) V(v)$ we get
\begin{equation}
\label{b-4}\frac{1}{V\,}\left(\varepsilon\frac{d^2 V}{dv^2}+\frac{dV}{dv}\right)=\frac{1}{U}\,\,\frac{d^2 U}{du^2}=-\alpha^2.
\end{equation}
By operating as before we get $\alpha=2 \pi m$, and the solution we are looking for is given by
\begin{equation}
\label{b-5}n(u,v)=K_0+\sum_{m=1}^{\infty}\left\{C_{1m}\,\exp(\mu_{1m} v)+C_{2m}\,\exp(\mu_{2m} v)\right\}\,\cos(2 m \pi u),
\end{equation}
where
\begin{eqnarray}
\label{b-6}\mu_{1m}&=&-\frac{1}{2 \varepsilon}\left\{1-\sqrt{1-(4 m \pi)^2 \varepsilon}\right\},\\
\label{b-7}\mu_{2m}&=&-\frac{1}{2 \varepsilon}\left\{1+\sqrt{1-(4 m \pi)^2 \varepsilon}\right\}.
\end{eqnarray}
The coefficients $C_{1m}$ and $C_{2m}$ have to be determined by means of the initial conditions (\ref{b-3}). Note that in the present case the characteristics exponents $\mu_{1m}$ and $\mu_{2m}$ became complex, and hence the relaxation is no longer a simple decreasing exponent, when $m>1/(4 \pi \sqrt{\varepsilon})$. As we will see in the following, this circumstance will change the relaxation of the initial distribution of particles. By means of expansion (\ref{b-5}) the initial conditions (\ref{b-3}) can be written as
\begin{eqnarray}
\label{b-8}n(u,0)&=&K_0+\sum_{m=1}^{\infty}\,\left(C_{1m}+C_{2m}\right)\cos(2 m \pi u),\\
\label{b-9}\left(\frac{\partial n}{\partial v}\right)_ {v=0}&=&\sum_{m=1}^{\infty}\left(\mu_{1m} C_{1m}+\mu_{2m} C_{2m}\right)\cos(2 m \pi u)=0,
\end{eqnarray}
from which it follows that $K_0$ is still given by Eq.(\ref{a-14}), and $C_{1m}$ and $C_{2m}$ can be expressed in terms of $K_m$ given by Eq.(\ref{a-15}) as
\begin{eqnarray}
\label{b-10}C_{1m}&=&\frac{\mu_{2m}}{\mu_{2m}-\mu_{1m}}\,K_m\\
\label{b-11}C_{2m}&=&-\frac{\mu_{1m}}{\mu_{2m}-\mu_{1m}}\,K_m.
\end{eqnarray}
The solution of the problem in the hyperbolic approximation is given by (\ref{b-5}) with the coefficients defined by means of Eq.s(\ref{b-10},\ref{b-11}), according to the initial distribution $n(u,0)$.

\section{Comparison of the predictions of the parabolic and hyperbolic approximations}
Our aim is now to compare the evolution of the initial distribution of particles when the diffusion phenomenon is described by means of the parabolic, (\ref{a-3}), or hyperbolic, (\ref{b-2}), equations.

\begin{figure}
 \includegraphics[width=1\textwidth]{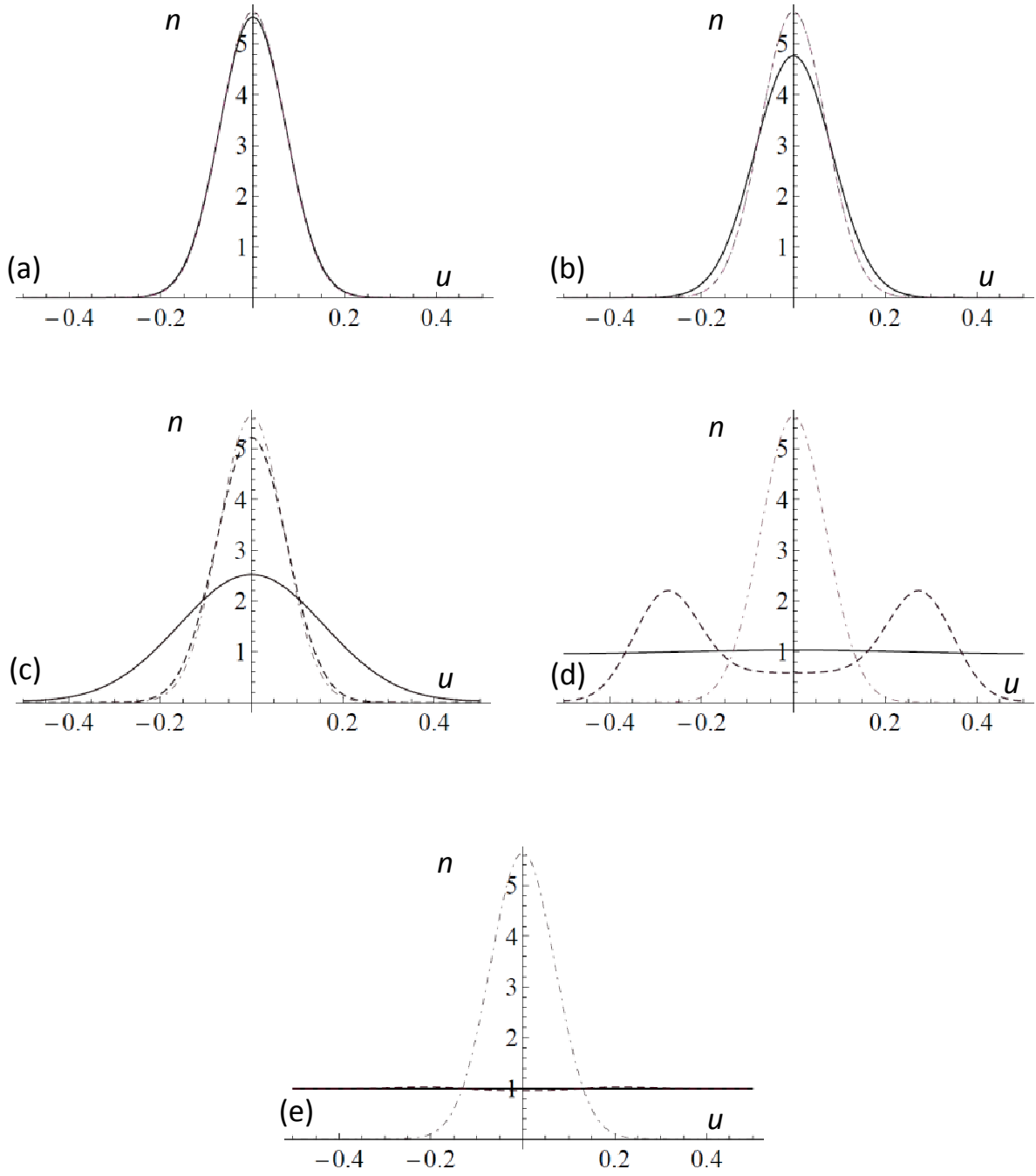}
\caption{ Evolution of the bulk density of particles when the initial distribution is assumed of the type $n(u,0)=B\sqrt{b/\pi}\,\exp(-b u^2)$, with $b=10^2$ and $B=1$ (dot-dashed curve), for different dimensionless time $v=t/\tau_D$, where $\tau_D=d^2/D$ is the diffusion time, and for $\varepsilon=\tau_r/\tau_D\sim$ 0.13. The thin and dashed curves refer to the parabolic and hyperbolic models, respectively. The values of the bulk density of particles are evaluated by considering 500 terms in the series representing $n(u,v)$ in each model. $v=10^{-4}$ (a), $v=10^{-3}$ (b), $v=10^{-2}$ (c), $v=10^{-1}$ (d), and $v=1$ (e).}
\label{fig1}       
\end{figure}

Let us consider first the situation where $n(u,0)$ is given by Eq.(\ref{a-16}). In Fig.\ref{fig1} we show the profile of density $n(u,v)$ across the sample for a few values of $v$. The dot-dashed curve represents the initial distribution of particles (\ref{a-16}), the thin curve the prediction of the parabolic approximation, and the dashed curve the prediction of the hyperbolic model. For small $v\ll1$, the three curves are practically coincident. As soon as $v$ increases, the evolution predicted by the parabolic model changes more rapidly than that of the hyperbolic model. However, the bulk density of the parabolic model tends to the equilibrium value in a monotonic manner, whereas in the case of the hyperbolic description not. This is specially evident from Fig.\ref{fig1}d. We note that for $v\sim 1$ the equilibrium distribution is reached, in the two approaches.

\begin{figure}
 \includegraphics[width=1\textwidth]{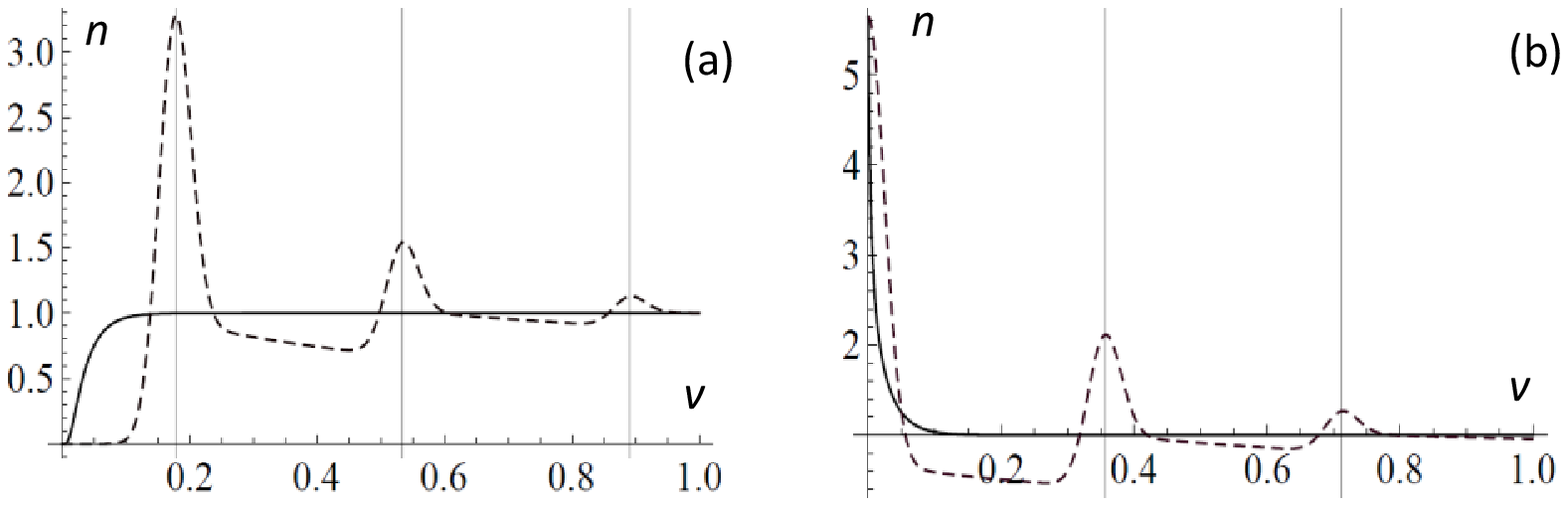}
\caption{ Time evolution of the bulk density of particles at a given point of the sample when the initial distribution is assumed of the type $n(u,0)=B\,\sqrt{b/\pi}\,\exp(-b u^2)$, with $b=10^2$ and $B=1$, for different dimensionless time $v=t/\tau_D$, where $\tau_D=d^2/D$ is the diffusion time, and for $\varepsilon=\tau_r/\tau_D\sim$ 0.13. Continuous line, parabolic model, dashed line hyperbolic model. (a) $u=0.5$ corresponds to the bulk density on the limiting surface. The vertical lines are drawn for $v_1=0.5/c$, $3 v_1$ and $5 v_1$. (b) $u=0$ corresponds to the bulk density in the middle of the sample. The vertical lines are drawn for $v_2=1/c=2 v_1$, $2 v_2=4 v_1$.}
\label{fig2}       
\end{figure}

In Fig.\ref{fig2} is reported, for a given spatial coordinate $u$, the time evolution of density of particles. In Fig.\ref{fig2}a $u=0.5$, and hence $n(0.5,v)$ represents the bulk density on the surface. The continuous curve represents the prediction of the parabolic model, whereas the dashed curve that of the hyperbolic model. As it is evident from Fig.2a, in the parabolic approximation the bulk density of particles on the surface tends the the equilibrium value $n_{\rm eq}$ in a monotonic manner, whereas, according to the hyperbolic approximation it presents a non monotonic trend. The vertical lines in Fig.\ref{fig2} have been drawn for multiple of $v_1=0.5/c$, where $c=1/\sqrt{\varepsilon}$. The quantity $c$ represents, in dimensionless form, the speed of the perturbation $C$ discussed in the introduction. Note that $n(0.5,v)$ presents maxima for $v_1$, $3 v_1$, $5 v_1$ and so on. This numerical result can be easily interpreted. In fact at $v=0$ the particles start to diffuse, and the first wave of density reaches the limiting surface after a time $v_1$. After that the wave is reflected from the limiting surface, and it reaches the opposite surface after a time $3 v_1$, it is reflected again. However, the initial wave reflected at $u=-0.5$ travels toward the surface at $u=0.5$, and it reaches it after a time $v_1+2 v_1=3 v_1$, and so on. Note that the maxima of different order are reduced for the presence of the linear term, responsible for the attenuation. The predicted time dependence of the density on the limiting surface is in agreement with the experimental observation reported in \cite{adamson}, and discussed in relation with the adsorption properties of the limiting surfaces \cite{luiz1,luiz2}. In Fig.\ref{fig2}b $u=0$, and hence $n(0,v)$ represents the bulk density in the middle of the sample. Also in this case the time variation of the density is not monotonic. In the same figure we have drawn vertical lines for $v_2=1/c=2 v_1$, $2 v_2=4v_1$ and so on. As in the previous case the maxima of the bulk density can be easily interpreted as related to the reflection of the wave of density on the limiting surfaces.

\begin{figure}
 \includegraphics[width=1\textwidth]{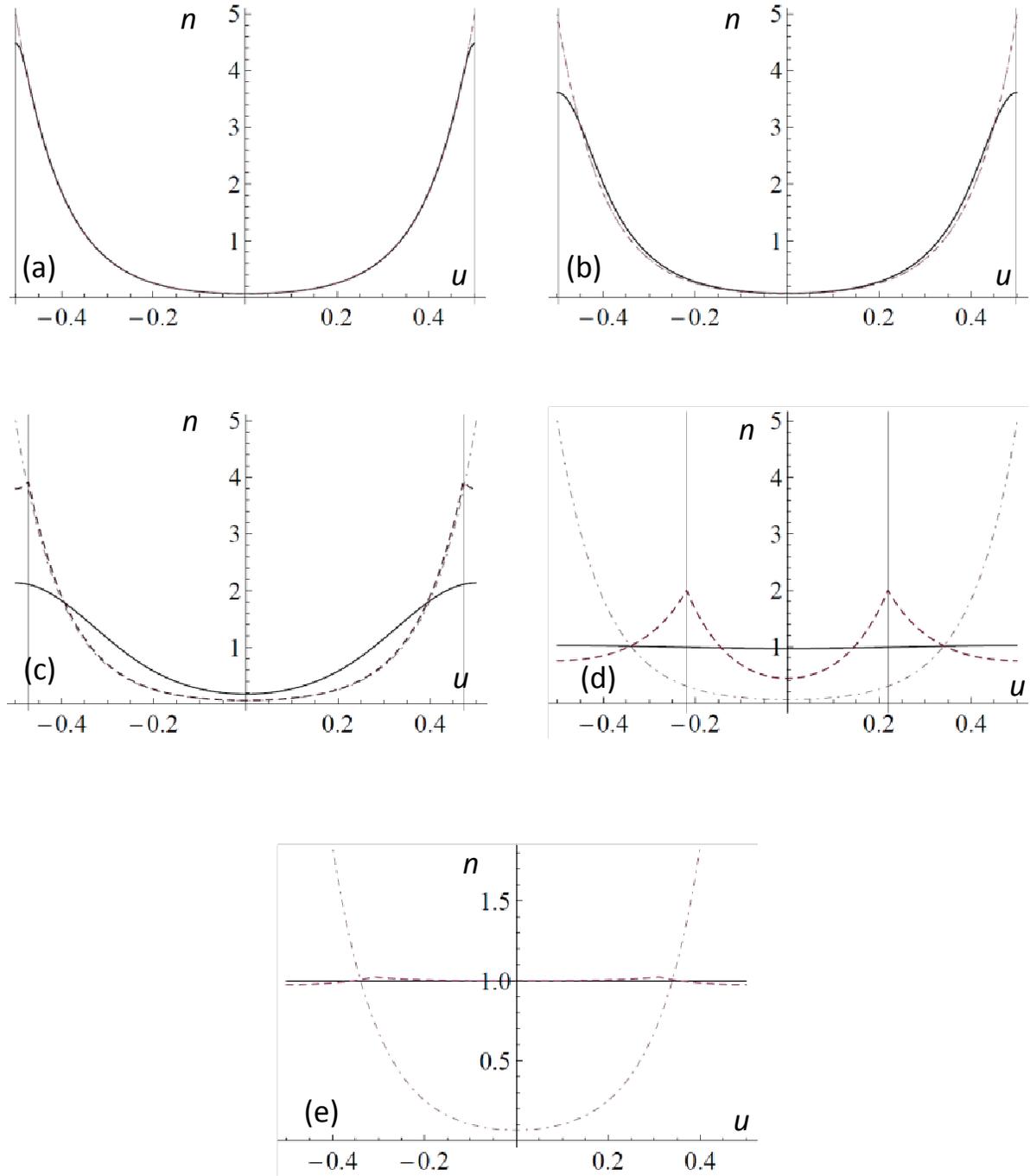}
\caption{ Evolution of the bulk density of particles when the initial distribution is assumed of the type $n(u,0)=\,A\,(s/2)\,\,\cosh(s u)/\sinh(s/2)$, with $s=10$ and $A=1$(dot-dashed curve), for different dimensionless time $v=t/\tau_D$, where $\tau_D=d^2/D$ is the diffusion time, and for $\varepsilon=\tau_r/\tau_D\sim$ 0.13. The thin and dashed curves refer to the parabolic and hyperbolic models, respectively. The values of the bulk density of particles are evaluated by considering 500 terms in the series representing $n(u,v)$ in each model. $v=10^{-4}$ (a), $v=10^{-3}$ (b), $v=10^{-2}$ (c), $v=10^{-1}$ (d), and $v=1$ (e). The vertical lines are drawn for $\pm u_1$, where $u_1=0.5-c v$.}
\label{fig3}       
\end{figure}

\begin{figure}
 \includegraphics[width=1\textwidth]{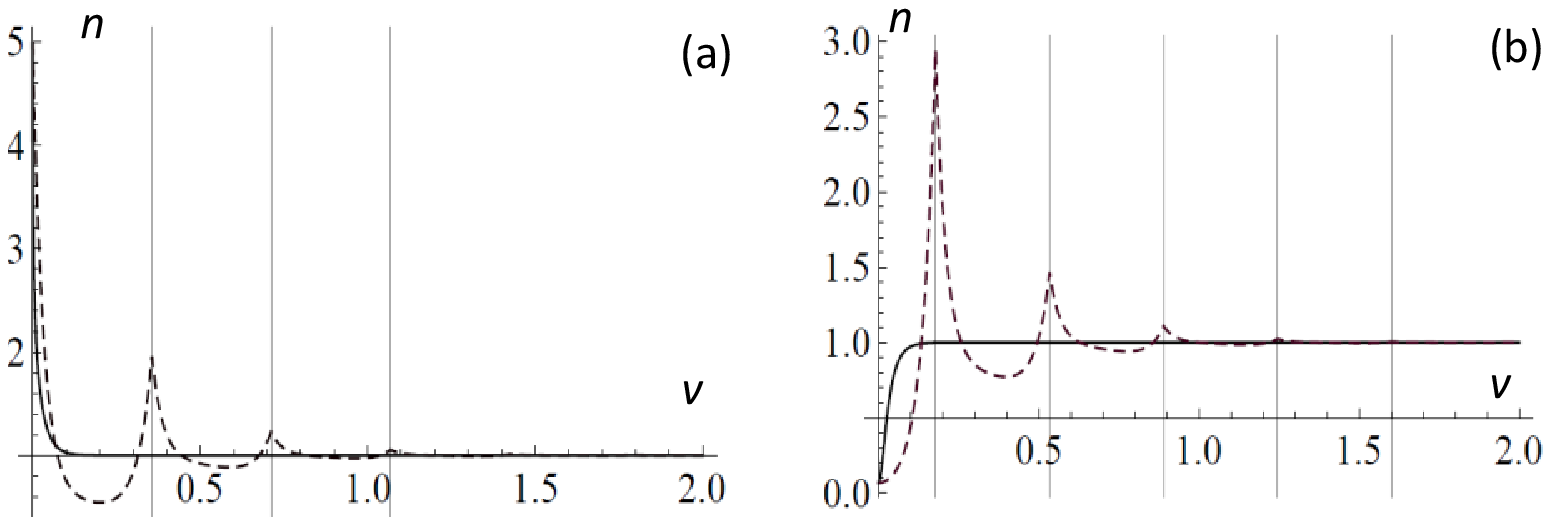}
\caption{  Time evolution of the bulk density of particles at a given point of the sample when the initial distribution is assumed of the type $n(u,0)=\,A\,(s/2)\,\,\cosh(s u)/\sinh(s/2)$, with $A=1$ and $s=10$, for different dimensionless time $v=t/\tau_D$, where $\tau_D=d^2/D$ is the diffusion time, and for $\varepsilon=\tau_r/\tau_D\sim$ 0.13. Continuous line, parabolic model, dashed line hyperbolic model. (a) $u=0.5$ corresponds to the bulk density on the limiting surface. The vertical lines are drawn for $v_2=1/c$, $2 v_2$ and $3 v_2$. (b) $u=0$ corresponds to the bulk density in the middle of the sample. The vertical lines are drawn for $v_1=0.5/c$, 3 $v_1$ and 5 $v_1$.}
\label{fig4}       
\end{figure}

Let us consider now the case where $n(u,0)$ is given by Eq.(\ref{a-19}). In Fig.\ref{fig3} we show, as in Fig.\ref{fig1}, the bulk density of particles across the sample for a few dimensionless times $v$. In this figure are also shown two vertical lines corresponding to $u_1=0.5-c v$, where there is a change of the function representing the distribution in the hyperbolic regime. Even in this case this result can be easily related to the fine speed of the wave of concentration. Finally, in Fig.\ref{fig4}, as in the previous Fig.\ref{fig2}, we show the bulk density of particles at the surface, $u=0.5$ (a), and in the bulk, $u=0$ (b). The vertical lines in (a) are drawn for $v_2=1/c$, $2 v_2$ and $3 v_2$, and in (b) for $v_1=0.5/c$, 3 $v_1$ and 5 $v_1$.

\section{WKB approach}
In the previous sections we have analysed the diffusion phenomenon in the parabolic and hyperbolic approximations.
We want now look for a method that is able to put out the
link between the two approaches. To this aim we use the WKB method \cite{5,6}. By operating as discussed above, from Eq.(\ref{b-4}) we get

\begin{equation}
  \varepsilon\,\frac{d^2 V}{dv^2}+\frac{d V}{dv} + \alpha^2 V = 0.
\label{eqditt1}
\end{equation}
We look for a solution of Eq.(\ref{eqditt1}) of the type $V(v)=V(v,\varepsilon)+B_{\alpha}\, e^{-\alpha^2 v}$ where
\begin{equation}
\label{x}V(v,\varepsilon)=e^{s(v)/{\varepsilon}},
\end{equation}
and the second contribution is the solution in the parabolic approximation, where $\varepsilon=0$.
By substituting this ansatz into Eq.(\ref{eqditt1}) we get
\begin{equation}
{\varepsilon} \left( \frac{s''(v)}{\varepsilon} +
\frac{(s'(v))^2}{\varepsilon^2} \right) +
\left( \frac{s'(v)}{\varepsilon} +  \alpha^2  \right) = 0 .
\label{hyp06}
\end{equation}
At the leading order $\varepsilon^{-1}$ from Eq.(\ref{hyp06}) we get
\begin{equation}
s'(v)[s'(v) + 1]= 0 ,
\label{hyp07}
\end{equation}
whose solutions are $ s(v) = -v/\varepsilon$ and and $s(v)$=constant. It follows that the correction to the parabolic solution is $V(v) = A\, {\rm exp}({-v/{\varepsilon}})$, where the constant is related to the part of solution of $s(v)$ which is $v$-independent.

Then the solution  $n(u,v)$ of Eq.(\ref{b-2}) by taking into account the boundary condition (\ref{a-2}) is of the type
\begin{equation}
n(u,v) =  \, K_0 \, + \, \sum_{m=1}^\infty
\left[  A_m \, e^{-v/{\varepsilon}}\, +
B_{m} \,e^{-(2 \pi m)^2 v}\right]  \cos (2m\pi u),
\label{hyp02bis}
\end{equation}
where the coefficients $  A_m$ and  $ B_{m}$ have to be determined by means
of the initial conditions on $n(u,v)$. A simple calculation gives
\begin{eqnarray}
\label{A}A_m&=&-\varepsilon\,\frac{(2 m \pi)^2}{1-\varepsilon(2 m \pi)^2}\,\,K_m,\\
\label{B}B_m&=&\,\,\frac{1}{1-\varepsilon(2 m \pi)^2}\,\,K_m.
\end{eqnarray}
Note that $\lim_{\varepsilon\to 0}A_m=0$ and $\lim_{\varepsilon\to 0}B_m=K_m$, as expected.

Equation (\ref{hyp02bis}) presents a term $A_m \, {\rm exp}({-v/{\varepsilon}} )$ that
disappears when $\varepsilon$ goes to $0^+$, but with $\varepsilon \neq 0$ gives a solution
that lies between the parabolic equation (\ref{a-3}) and the hyperbolic one (\ref{b-2}).

\section{Conclusions}
We have investigated the evolution of an initial distribution of diffusing particles in an isotropic medium in the parabolic and hyperbolic regimes. The sample has been assumed in the shape of a slab, and the initial distribution represented by an even function of the normal coordinate to the limiting surfaces. The simple case where the bounding surfaces are perfectly blocking has been considered. In this framework, as expected, in the parabolic regime the initial distribution tends to that of equilibrium in a monotonic manner, in the sense that in a given point of the sample the density of particles changes in a monotonic manner with time. On the contrary, in the hyperbolic regime, the distribution tends to that of equilibrium oscillating around the value of equilibrium.

\section*{Acknowledgements}
The research leading to these results has received funding from
the European Research Council under the European Union's Seventh
Framework Programme (FP/2007-2013)/ERC Grant Agreement
No. 306622 (ERC Starting Grant Multi-field and Multi-scale Computational
Approach to Design and Durability of PhotoVoltaic Modules-CA2PVM). The support of the Italian Ministry of Education,
University and Research to the Project FIRB 2010 Future in Research
Structural Mechanics Models for Renewable Energy Applications
(RBFR107AKG) is gratefully acknowledged.

\end{document}